\documentstyle[aps,pre,manuscript]{revtex}

\begin{document}

\draft

\title{Correlated noise induced control of prey extinction}
\author{Suman Kumar Banik {\footnote {e-mail: pcskb@mahendra.iacs.res.in} }
}
\address{Department of Physical Chemistry \\
Indian Association for the Cultivation of Science \\
Jadavpur, Calcutta 700 032, India}

\date{\today}

\maketitle

\begin{abstract}
We study the steady state properties of a phenomenological two-state 
predator model in presence of correlated Gaussian white noise. 
Based on the corresponding Fokker-Planck equation for probability 
distribution function the steady state solution of the probability 
distribution function and its extrema have been investigated. We 
show for a typical value of noise correlation there is a 
giant loss of bistability which in turn prevents the prey population
from going into extinction.
\end{abstract}


\section{Introduction}
The subject of noise-induced transition has got wide applications in
the field of physics, chemistry and biology \cite{hl}. In most of these 
theories the noise affects the dynamics through system variable, i.e., the
noise is multiplicative in nature. The focal theme of these investigations 
is the study of steady state properties of the system where the
fluctuations, in general, are applied from outside and are independent
of the system's characteristic dissipation. Such systems are generally
termed as open systems \cite{lw}, since they lack the principle of detailed
balance which ensures fluctuation-dissipation relation to hold for the
thermodynamically closed systems. However, it may also happen that the
external fluctuations instead of affecting only some system's parameters 
affect the system directly, i.e., they drive system dynamics multiplicatively 
as well as additively. Because the two noise processes owe a common origin 
they get correlated in the relevant timescale of the problem \cite{ft,cw}. 
Correlated noise processes have found applications in studying steady
state properties of a single mode laser \cite{szhu}, in
analyzing bistable kinetics \cite{wck}, in giant supression of activation
rate \cite{ajrm}, in producing
directed motion in spatially symmetric periodic potentials \cite{lh}, 
in studying stochastic resonance in linear systems \cite{bg}, 
in steady state entropy production \cite{bbr}, etc.
In this brief communication we investigate a simple noise-driven two-state
predator model \cite{hl} and show how noise correlation can dynamically 
prevent the prey population from extinction.

\section{The Model}

To start with we consider an environment of the prey which in absence
of predation grows logistically and at the same time its density in a
territory depends linearly on a constant source of migration. We also 
consider a population of predators in the given territory which lives by 
feeding on prey. The characteristic time scale over which the population
of prey and predator varies are very much different, so one can consider
the predator population to be constant within the generation time of prey.
The predators are engaged in two types of activities, viz, hunting or resting.
The time scale of predator's two activities are very short compared to the
generation time of prey, i.e., $\tau_R$, $\tau_H$ $\ll$ $\mu^{-1}$ where
$\tau_R$ and $\tau_H$ are the characteristic average time of resting and
hunting, respectively and $\mu$ is the birth rate of prey. 
The activity of the predator in the territory 
resembles the mode of action of enzymes or catalysts in a chemical reaction.
The enzymes or catalysts in a chemical reaction transform substrates in
a continuous manner without destroying themselves. The constant
predator population acts in a similar way by feeding on the prey. To put this 
ideas in a quantitative way we write the evolution equations for the
predator and prey \cite{hl},
\begin{eqnarray}
\label{eq1}
\dot{X} & = & A + \mu X \left ( 1 - \frac{X}{K} \right ) - 
\frac{1}{\tau_H} XY \; \; , \\
\label{eq2}
\dot{Y} & = & - \; \frac{1}{\tau_H} XY + \frac{1}{\tau_R} Z
\end{eqnarray}

\noindent
where $X$ is the density of prey in a given territory. The constant $A$
in Eq.(\ref{eq1}) is due to a constant source of prey through immigration. 
The second term in (\ref{eq1}) is the Fisher logistic growth term with birth rate 
$\mu$ and carrying capacity $K$. $Y$ and $Z$ are the numbers of predators in 
the hunting and resting state, respectively. $E$ is the total constant 
population of the predators, i.e., $E \equiv Y(t) + Z(t) =$ constant.
The last term in (\ref{eq1}) describes the decay rate of prey. The model
is hybrid in nature in the sense that it has virtue of taking into consideration
of the logistic growth model as well as of the predator-prey model.

Following Ref.\cite{hl}
we now consider that the predator population, $E$ is small compared to
prey population $X$. To study the overall dynamics within the timescale 
$\mu^{-1}$ we make the following transformation 
\begin{equation}
\label{eq3}
\tau_H = \varepsilon \tau_H^* \; \; , \; \; \tau_R = \varepsilon \tau_R^*
\; \; , \; \; Y = \varepsilon Y^*  \; \; {\rm and} \; \;
Z = \varepsilon Z^*
\end{equation}

\noindent
where $\varepsilon$ is a small quantity, $\tau_H^*$, $\tau_R^*$ are 
quantities of order $\mu^{-1}$ and $Y^*$, $Z^*$ are quantities of order $X$. 
Using (\ref{eq3}) in (\ref{eq1}) and (\ref{eq2}) we arrive at
\begin{eqnarray}
\label{eq4}
\dot{X} & = & A + \mu X \left ( 1 - \frac{X}{K} \right ) - 
\frac{1}{\tau_H^*} XY^* \; \; , \\
\label{eq5}
\varepsilon \dot{Y}^* & = & - \; \frac{1}{\tau_H^*} XY^* + 
\frac{1}{\tau_R^*} Z^* \; \; .
\end{eqnarray}

Now eliminating $Y^*$ from (\ref{eq4}) and using the limit 
$\varepsilon \rightarrow 0$ we arrive at the following dimensionless
evolution equation for prey
\begin{equation}
\label{eq6}
\dot{x} = \alpha + x (1-\theta x) - \beta \frac{x}{1+x}
\end{equation}

\noindent
where
\begin{equation}
\label{eq7}
x = \frac{\tau_R^*}{\tau_H^*} X \; \; , \; \; 
\alpha = \frac{A \tau_R^*}{ \mu \tau_H^*} \; \; , \; \;
\beta = \frac{E}{\mu \tau_H} \; \; {\rm and} \; \;
\theta = \frac{\tau_H}{\tau_R K} \; \; .
\end{equation}

\noindent
It is interesting to note that the third term in Eq.(\ref{eq6}) is the
predation term which essentially emerges from the two-state of predator
activities. The steady state solution of Eq.(\ref{eq6}) shows a cusp
type of catastrophe. The corresponding critical point 
($\alpha_c, \beta_c, x_c$) is given by \cite{hl}
\begin{eqnarray*}
\alpha_c = \frac{(1-\theta)^2}{27\theta^2} \; \; , \; \;
\beta_c = \frac{(1+2\theta)^3}{27\theta^2} \; \; {\rm and} \; \;
x_c = \frac{1-\theta}{3\theta} \; \; .
\end{eqnarray*}

\noindent
The necessary condition to have a physically realizable critical point
i.e., for $\alpha_c$, $x_c$ to be positive, is $\theta < 1$. Thus the 
steady state curve of $x$ as a function of $\beta$ always shows a bistable
region for small values of $\theta$. The smallness condition may be
maintained by increasing the carrying capacity $K$ or by decreasing
the ratio $\tau_H/\tau_R$.

Eq.(\ref{eq6}) is the starting point of our further analysis. It may be noted
that $\alpha$ and $\beta$ are the two quantities which appear in the
prey evolution equation as a constant and a multiplicative factor,
respectively. Expressions for $\alpha$ and $\beta$ in (\ref{eq7}) suggest
that they are connected by a common parameter $\mu$, the birth rate
of the prey. Now if due to some environmental external disturbance the birth
rate of the prey fluctuates, it is likely to affect both $\alpha$ and
$\beta$ in the form of additive and multiplicative noises which are
connected through a correlation parameter. Or in other words the
external fluctuations affect the parameter
$\beta$ which fluctuates around a mean value, thus generating 
multiplicative noise and at the same time environmental fluctuations 
perturbs the dynamics directly
which gives rise to additive noise. As a result we have the stochastic
differential equation in Stratonovich prescription,
\begin{equation}
\label{eq8}
\dot{x} = \alpha + x (1-\theta x) - \beta \frac{x}{1+x} -
\frac{x}{1+x} \xi (t) + \eta (t)
\end{equation}

\noindent
where $\xi (t)$ and $\eta (t)$ are the stationary Gaussian white noises
with the following properties

\begin{eqnarray}
\langle \xi (t) \rangle & = & \langle \eta (t) \rangle = 0  \; \; , \\
\langle \xi (t) \xi (t') \rangle & = & 2 \sigma \delta (t-t') \; \; ,
\\
\langle \eta (t) \eta (t') \rangle & = & 2 D \delta (t-t') \; \; {\rm and}
\\
\langle \xi (t) \eta (t') \rangle & = & \langle \eta (t) \xi (t') \rangle
= 2 \lambda (\sigma D)^{1/2} \delta (t-t')
\end{eqnarray}


\noindent
where $\lambda$ denotes the degree of correlation between noise processes
$\xi (t)$ and $\eta (t)$
with $ 0 \leq \lambda \leq 1$. Using the above mentioned noise properties
we write down the corresponding Fokker-Planck equation (in Stratonovich 
prescription) for the evolution of probability distribution function 
\cite{cw,wck},
\begin{equation}
\label{eq9}
\frac{\partial}{\partial t} P(x,t) = - \frac{\partial}{\partial x}
A(x,t) P(x,t) + \frac{\partial^2}{\partial x^2} B (x,t) P(x,t)
\end{equation}

\noindent
where

\begin{equation}
A (x,t) = 
\alpha + x (1-\theta x) - \beta \frac{x}{1+x} +
\sigma \frac{x}{(1+x)^3}  - \lambda (\sigma D)^{1/2} \frac{1}{(1+x)^2} 
\end{equation}
and
\begin{equation}
B (x,t)  =  D + \sigma \frac{x^2}{(1+x)^2}  - 
2 \lambda (\sigma D)^{1/2} \frac{x}{1+x} \; \; .
\end{equation}


\section{Steady state analysis and results}

Using the zero current condition at the stationary state we derive
the stationary probability distribution function (SPDF) with $0$ and 
$\infty$ as the natural boundaries,
\begin{equation}
P_s (x) = N \frac{1}{ B(x)} \exp \left [ 
\int^x \frac{ A(x') }{ B(x') } dx' \right ] 
\end{equation}

\noindent
where $N$ is the normalization constant. Using the explicit
forms of $A(x)$ and $B(x)$ we have the following explicit forms of
SPDF
\begin{equation}
P_s (x) = N (1+x) g^{\nu-\frac{1}{2}} (x)
\exp [ q_1 x^3 + q_2 x^2 + q_3 x + q_4 f(x) ]
\end{equation}

\noindent
where
\begin{equation}
g(x) = a + bx + cx^2
\end{equation}

\begin{equation}
\label{eq10}
\begin{array}{ccccc}
f (x) & = & -2/(b+2 c x) &  {\rm for} & \lambda = 1 \\
      & = & (2/ \sqrt{\Delta} ) \arctan [ (b+2cx)/ \sqrt{\Delta} ]
      &  {\rm for} &  0 \leq \lambda < 1
\end{array}
\end{equation}

\noindent
with
\begin{eqnarray}
\label{eq11}
a & = &  D \; , \;  b = 2 [ D-\lambda (\sigma D)^{1/2} ] \; ,
\nonumber \\
c & = & D + \sigma - 2 \lambda (\sigma D)^{1/2} \; , 
\Delta = 4 \sigma D ( 1 - \lambda^2 )
\end{eqnarray}

\noindent
along with
\begin{eqnarray}
q_1 & = & \frac{-\theta}{3c} \; , \nonumber \\
q_2 & = & \frac{1-2\theta}{2c} + \frac{b \theta}{2 c^2} \; , \;
q_3 =  \frac{\alpha -\beta - \theta +2}{c} - 
\frac{b (1-2\theta) + a \theta}{c^2} - \frac{b^2 \theta }{c^3} \; , 
\nonumber \\
q_4 & = & \alpha - \frac{ b (2\alpha-\beta+1)}{2c} +
\frac{(b^2-2ac)(\alpha-\beta-\theta+2)}{2c^2} - 
\frac{ b^2(b^2-3ac) \theta}{2c^4} \nonumber \\
& &  + \frac{ a(b^2-2ac) \theta - b (b^2-3ac) (1-2\theta)}{2c^3}  \; \;
{\rm and} \nonumber \\
\nu & = & \frac{2\alpha-\beta+1}{2c} -
\frac{ b ( \alpha -\beta \theta +2)}{2c^2} +
\frac{ (b^2-ac)(1-2\theta) - ab\theta}{2c^3} \nonumber \\
& & + \frac{b (b^2-ac) \theta}{2c^4} \; \; .
\end{eqnarray}


\noindent
The extrema of SPDF is calculated using the condition $A(x)-B'(x)=0$,
\begin{equation}
\label{eq12}
\alpha + x (1-\theta x) - \frac{\beta x}{1+x} - 
\frac{\sigma x}{(1+x)^3} + \frac{\lambda (\sigma D)^{1/2} }{ (1+x)^2} = 0
\; \;  {\rm for} \; 0 \leq \lambda \leq 1 \; \; .
\end{equation}

\noindent
For zero noise correlation, i.e., for $\lambda=0$ the last term of 
Eq.(\ref{eq12}) vanishes and we have the extrema of SPDF for pure 
multiplicative noise processes \cite{hl}. For zero correlation the additive 
noise has no extra effects in the steady state dynamics. To illustrate 
this we have plotted extrema of SPDF as a function of $\beta$ in 
Fig.(1) using the parameters given in \cite{hl}. For zero noise correlation 
the curve shows a sharp minima which decreases on increasing $\lambda$. 
Similarly, in Fig.(2) we have plotted extrema of SPDF as a 
function of $\beta$ for different values of additive noise strength $D$ 
with maximum correlation ($\lambda=1$). As the additive noise strength 
increases the well gets flattened and almost vanishes for a large enough 
value of $D$.

In Fig.(3) we show the effect of correlation parameter $\lambda$ on SPDF. 
For a low value of $\lambda$ the SPDF shows the typical bistable region 
( see Fig.3(a) ) which vanishes for higher values of $\lambda$ ( see 
Fig.3(b) ). As the value of correlation parameter 
$\lambda$ increases the peak on the lower values of $x$ decreases while
for a higher value of $\lambda$ we have a single peak at a higher
values of $x$. Since $x$ denotes the prey population, it is clear from
Fig.(3) that with the increase of $\lambda$ values the prey population 
recovers from going into extinction. In other words, the distribution
of prey which was mainly peaked about zero (for a low value of $\lambda$)
signifying high extinction rate, moves away from zero with the increase of 
correlation between noises thus favouring the prey's survival.
Though Gaussian white noise acting independently and
multiplicatively favours the extinction of prey \cite{hl}, the
extinction rate decreases drastically for a simultaneous perturbation
of additive and multiplicative white noise originating from a common source,
hence connected through a correlation parameter.

From the expressions of $f(x)$, $b$ and $c$ given in Eqs.(\ref{eq10}) and
(\ref{eq11}) it is clear that for $\lambda$ = 1.0 we have always a singular
distribution for $\sigma=D$, since it makes both the parameter $b$ and $c$
zero and eventually leads to the divergence of all the  $q$'s and $\nu$.
However this divergence can be removed for appreciable difference between
the $\sigma$ and $D$ values. In Fig.(4) we have plotted the typical behaviour 
of SPDF for maximum correlation $\lambda$ = 1 which shows monotonic
decreasing behaviour. In contrast to the behaviour shown in Fig.(3), Fig.(4),
however, shows the hastening of prey's extinction for a full correlation
between additive and multiplicative noises.

In this brief communication we have studied the effect of environmental 
fluctuation of the birth rate of the prey in terms of external correlated 
noise processes which appreciably modify the macroscopic behaviour of a 
two-state predator model. We have shown how the correlation between the two 
noise processes which owe a common origin may drastically prevent the
extinction of the prey.

I express my sincerest gratitude to Prof. D S Ray for suggesting me the 
problem and for his continuous inspiration during the progress of this work.
This work was supported by Council of Scientific and Industrial Research
(C.S.I.R.), Govt. of India.

\begin{figure}[h]
\caption{ Plot of extrema of SPDF
as a function of $\beta$ for different values of noise
correlation $\lambda$ using $\alpha$ = 4.5, $\theta$ = 0.1, $\sigma$ = 33.0
and $D$ = 3.0.
}
\label{fig1}
\end{figure}

\begin{figure}[h]
\caption{ Same as in Fig.(1) but for different values of additive noise
strength $D$. The other parameters are same except $\lambda$ = 1.
}
\end{figure}

\begin{figure}[h]
\caption{ Plot of $P_s(x)$ against $x$ for different values of noise
correlation $\lambda$ using $\alpha$ = 4.5, $\theta$ = 0.1, $\beta$ = 7.5,
$\sigma$ = 3.0 and $D$ = 0.3.
(a) For low values of $\lambda$ and (b) for high values of $\lambda$.
}
\end{figure}

\begin{figure}[h]
\caption{ Same as in Fig.(3) but for $\lambda$ = 1.0 and $D$ = 2.12.
}
\end{figure}


\begin{thebibliography}{99}

\bibitem{hl} W. Horsthemke and R. Lefever, {\it Noise-Induced Transitions}
(Springer-Verlag, Berlin, 1984).

\bibitem{lw} K. Lindenberg and B. West, {\it The Nonequilibrium Statistical
Mechanics of Open and Closed Systems} (VCH, New York, 1990).

\bibitem{ft} A. Fulinski and T. Telejko, Phys. Lett. A {\bf 152}, 19 (1991).

\bibitem{cw} Li Cao and Da-jin Wu, Phys. Lett. A {\bf 185}, 59 (1994).

\bibitem{szhu} S. Zhu, Phys. Rev. A {\bf 47}, 2405 (1993).

\bibitem{wck} Wu Da-jin, Cao Li and Ke Sheng-zhi, Phys. Rev. E {\bf 50},
2496 (1994); Ya Jia and Jia-rong Li, Phys. Rev. E {\bf 53}, 5786 (1996).

\bibitem{ajrm} A. J. R. Madureira, P. H\"anggi and H. S. Wio, Phys. Lett. A
{\bf 217}, 248 (1996).

\bibitem{lh} J. H. Li and Z. Q. Huang, Phys. Rev. E {\bf 53}, 3315 (1996);
{\it ibid} {\bf 57}, 3917 (1998).

\bibitem{bg} V. Berdichevsky and M. Gitterman, Phys. Rev. E {\bf 60},
1494 (1999)

\bibitem{bbr} B. C. Bag, S. K. Banik and D. S. Ray, Phys. Rev. E {\bf 64},
026110 (2001).


\end{thebibliography}
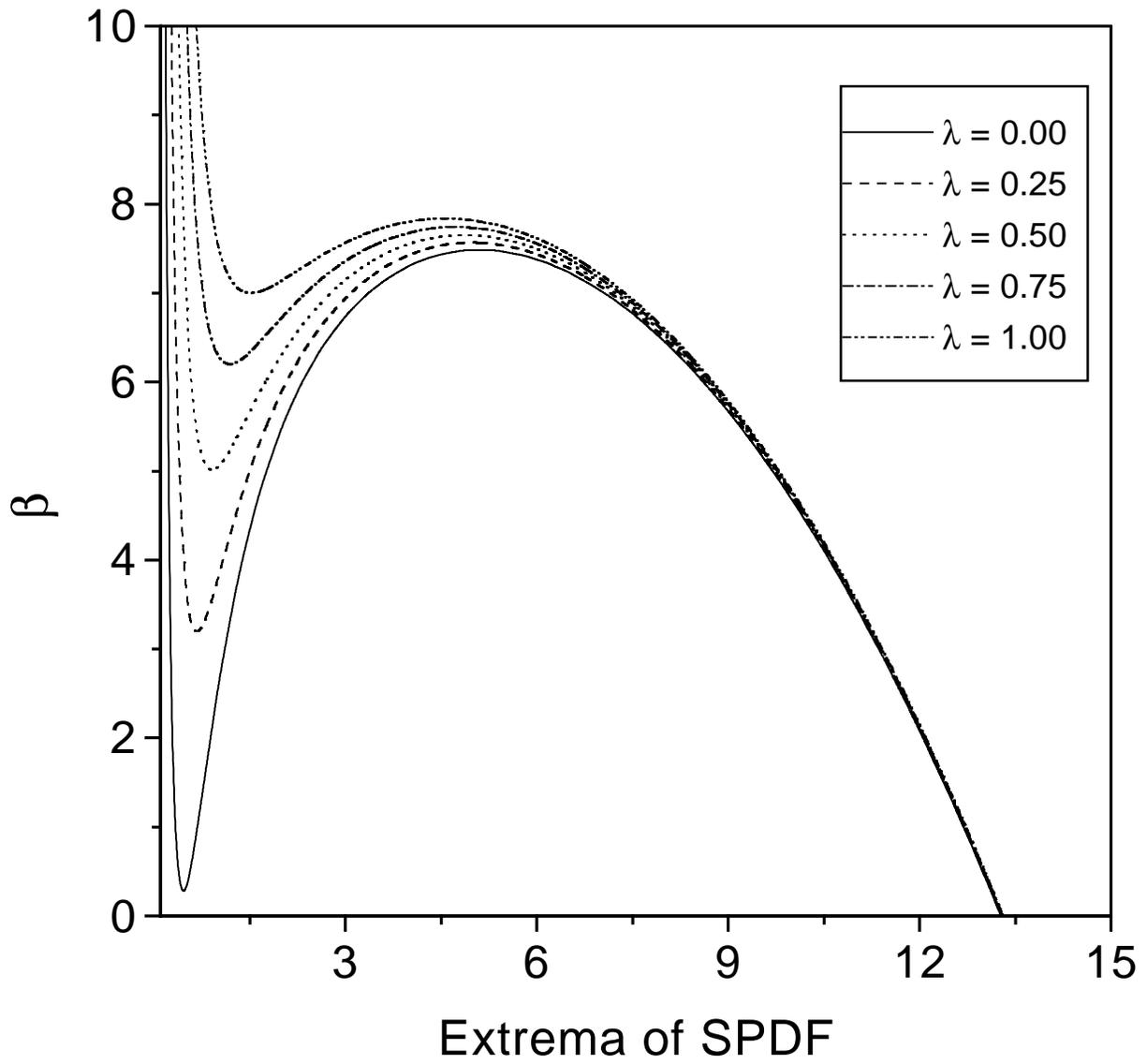
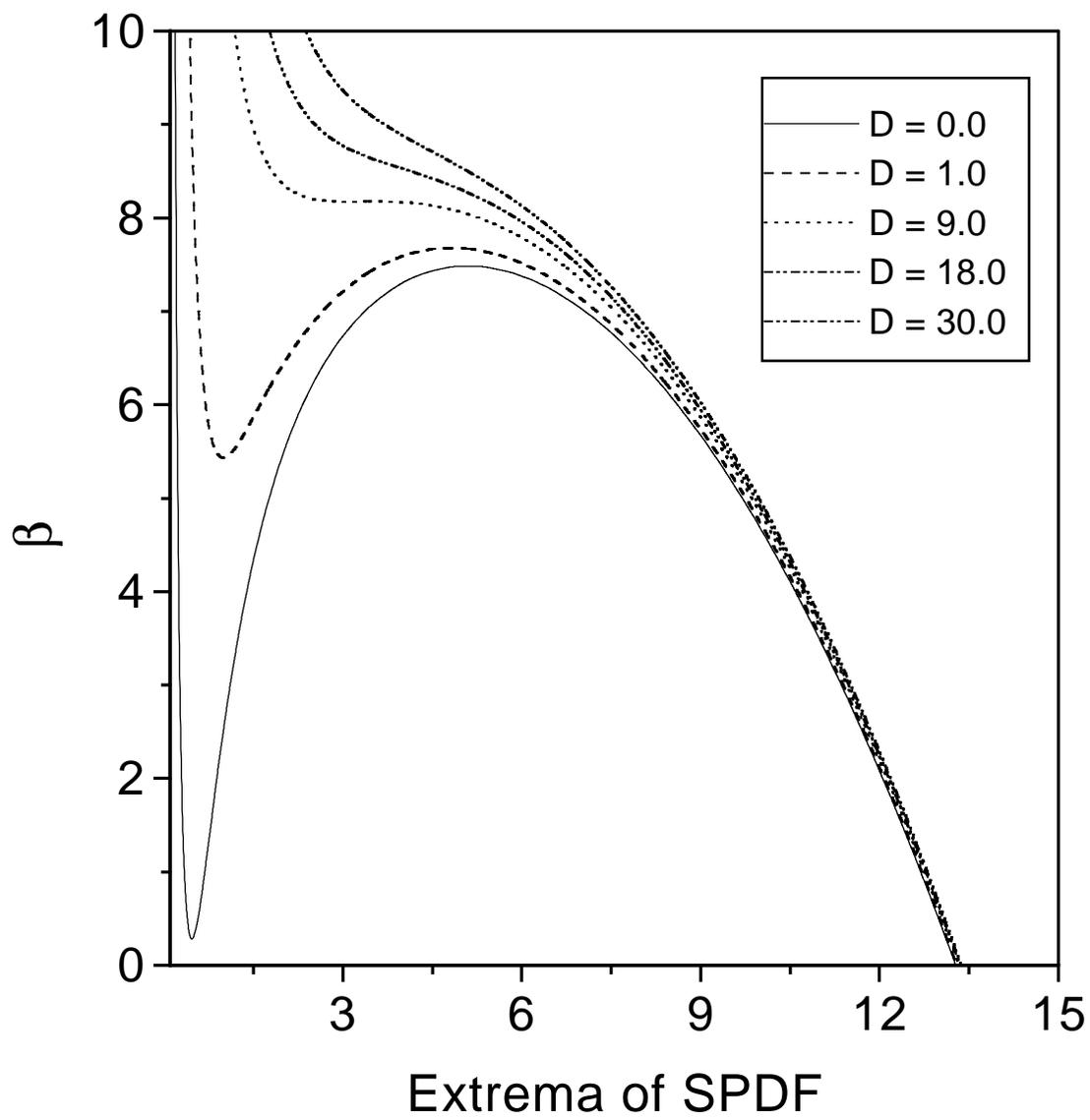
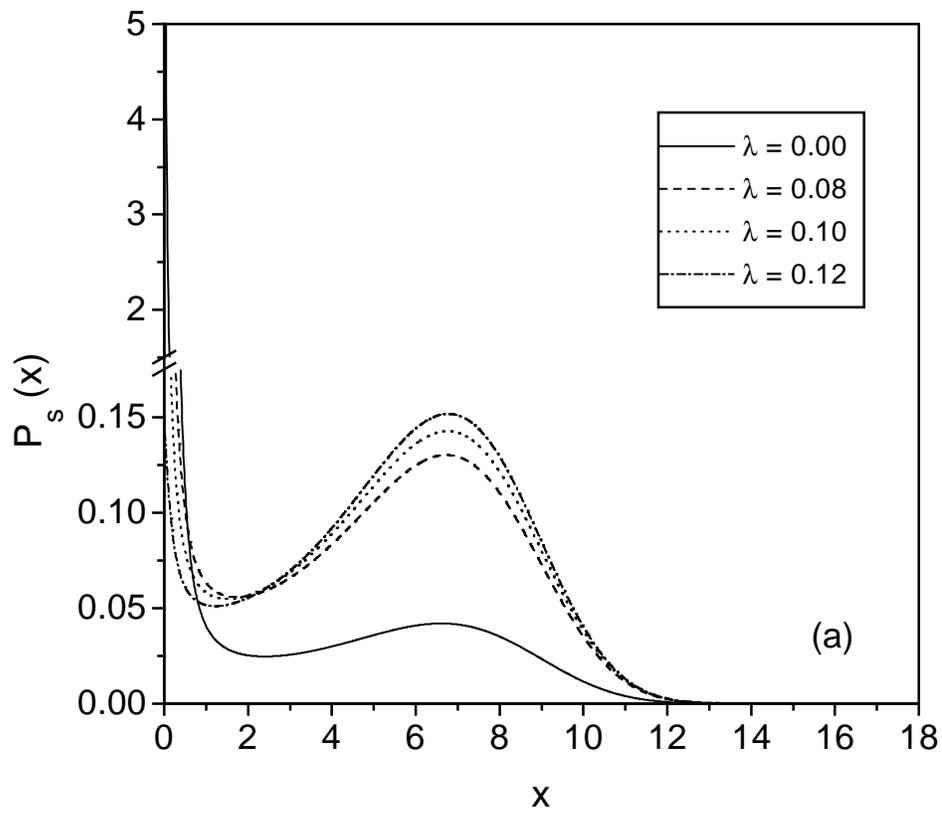
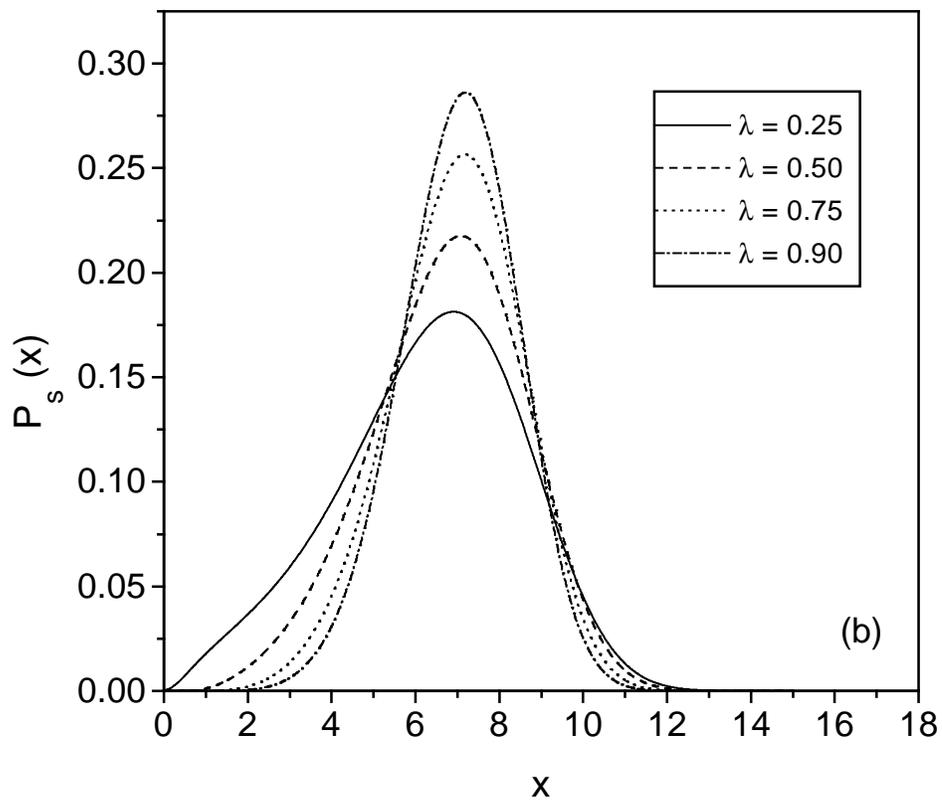
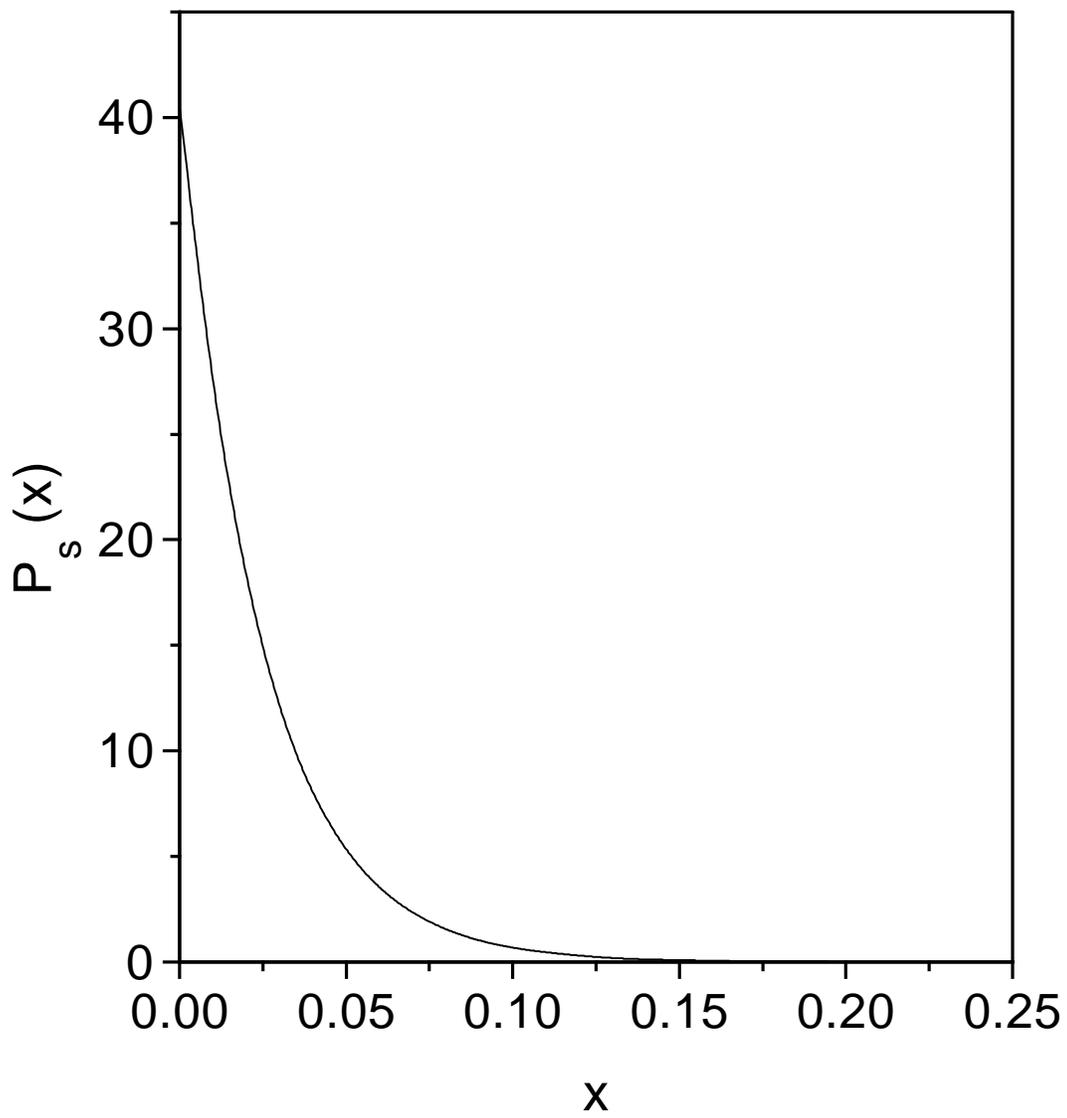
\end{document}